\pdfoutput=1

\documentclass[sigconf,screen=true,natbib,9pt]{acmart}

\copyrightyear{2018}
\acmYear{2018}
\setcopyright{none}
\acmConference[DLRS 2018]{DLRS 2018}{October 6, 2018}{Vancouver, Canada}

\usepackage{geometry}
\usepackage{dsfont}
\usepackage{xAlgorithm}
\usepackage{booktabs}
\usepackage{paralist}
\usepackage{amsmath}
\usepackage{graphicx}
\usepackage{subfigure}
\usepackage{xspace}
\usepackage{multirow}
\usepackage{color}
\usepackage[skip=0pt]{caption}
\usepackage{enumitem}
\usepackage{stmaryrd}
\usepackage{amssymb}
\usepackage[np, autolanguage]{numprint}

\usepackage{pgf,tikz}
\usetikzlibrary{arrows,arrows.meta,shapes,matrix,fit,backgrounds,decorations.shapes,graphs,quotes,patterns,decorations.pathreplacing}

\usepackage{acronym}
\acrodef{VAE}{Variational Autoencoder}
\acrodef{cVAE}{collective Variational Autoencoder}
\acrodef{CF}{Collaborative Filtering}
\acrodef{MF}{Matrix Factorization}
\acrodef{SLIM}{Sparse LInear Method}
\acrodef{cSLIM}{collective SLIM}
\acrodef{rcSLIM}{relex cSLIM}
\acrodef{fSLIM}{Side Information Induced SLIM}
\acrodef{MLP}{Multi-Layer Perceptron}
\acrodef{ELBO}{Evidence Lower Bound}
\acrodef{UAE}{User-side Autoencoder}
\acrodef{IAE}{Item-side Autoencoder}
\acrodef{FAE}{Feature-side Autoencoder}
\acrodef{Rec@N}{Recall at $N$}
\acrodef{Pre@N}{Precision at $N$}
\acrodef{MAP@N}{Mean average precision at $N$}
\acrodef{AP@N}{Average precision at $N$}

\newcommand{\set}[1]{\left\{#1\right\}}

\renewcommand{\vec}[1]{\boldsymbol{#1}}
\newcommand{\size}[1]{\left|#1\right|}

\newcommand{\rbr}[1]{\left(#1\right)}
\newcommand{\sbr}[1]{\left[#1\right]}

\newcommand{\cslim}{\textsf{cSLIM}\xspace}

\newcommand{\cvae}{\textsf{cVAE}\xspace}
\newcommand{\cfvae}{\textsf{cfVAE}\xspace}
\newcommand{\rvae}{\textsf{rVAE}\xspace}
\newcommand{\fvae}{\textsf{fVAE}\xspace}

\newcommand{\sport}{\textsf{Sports}\xspace}
\newcommand{\game}{\textsf{Games}\xspace}

\newcommand{\goodgap}{\hspace{\subfigtopskip}\hspace{\subfigbottomskip}}

\allowdisplaybreaks

\title[A Collective Variational Autoencoder]{A Collective Variational Autoencoder for Top-$N$ \\ Recommendation with Side Information}

\author{Yifan Chen}
\affiliation{%
  \institution{University of Amsterdam}
}
\email{y.chen4@uva.nl}

\author{Maarten de Rijke}
\orcid{0000-0002-1086-0202}
\affiliation{%
\institution{University of Amsterdam}
}
\email{derijke@uva.nl}

\begin{document}


\begin{abstract}
Recommender systems have been studied extensively due to their practical use in many real-world scenarios.
Despite this, generating effective recommendations with sparse user ratings remains a challenge.
Side information associated with items has been widely utilized to address rating sparsity.
Existing recommendation models that use side information are linear and, hence, have restricted expressiveness.
Deep learning has been used to capture non-linearities by learning deep item representations from side information but as side information is high-dimensional existing deep models tend to have large input dimensionality, which dominates their overall size.
This makes them difficult to train, especially with small numbers of inputs.

Rather than learning item representations, which is problematic with high-dimensional side information, in this paper, we propose to learn feature representation through deep learning from side information.
Learning feature representations, on the other hand, ensures a sufficient number of inputs to train a deep network.
To achieve this, we propose to simultaneously recover user ratings and side information, by using a \ac{VAE}. 
Specifically, user ratings and side information are encoded and decoded collectively through the same inference network and generation network.
This is possible as both user ratings and side information are data associated with items.  
To account for the heterogeneity of user rating and side information, the final layer of the generation network follows different distributions depending on the type of information.
The proposed model is easy to implement and efficient to optimize and is shown to outperform state-of-the-art top-$N$ recommendation methods that use side information.
\end{abstract}

%
%



\keywords{Top-$N$ recommendation, Side information, Collective Variational autoencoder}

\maketitle


\section{Introduction}
\label{section:introduction}
Recommender systems have become increasingly indispensable.
Applications include top-$N$ recommendations, which are widely adopted to recommend users ranked lists of items.
For e-commerce, typically only a few recommendations are shown to the user each time and recommender systems are often evaluated based on the performance of the top-$N$ recommendations.

\ac{CF} based methods are a fundamental building block in many recommender systems. \ac{CF} based recommender systems predict what items a user will prefer by discovering and exploiting similarity patterns across users and items.
The performance of \ac{CF}-based methods often drops significantly when ratings are very sparse. 
With the increased availability of so-called \emph{side information}, that is, additional information associated with items such as product reviews, movie plots, etc., there is great interest in taking advantage of such information so as to compensate for the sparsity of ratings.

Existing methods utilizing side information are linear models~\cite{DBLP:conf/recsys/NingK12}, which have a restricted model capacity. 
A growing body of work generalizes linear model by deep learning to explore non-linearities for large-scale recommendations~\cite{DBLP:conf/www/HeLZNHC17,DBLP:conf/www/SedhainMSX15,DBLP:conf/wsdm/WuDZE16,DBLP:conf/icml/ZhengTDZ16}.
State-of-the-art performance is achieved by applying \acfp{VAE}~\cite{DBLP:journals/corr/KingmaW13} for \ac{CF}~\cite{DBLP:conf/kdd/LiS17,DBLP:conf/www/LiangKHJ18,DBLP:conf/cikm/LeeSM17}.
These deep models learn item representations from side information.
Thus, the dimension of side information determines the input dimension of the network, which dominates the overall size of the model.
This is problematic since side information is generally high-dimensional~\cite{DBLP:conf/sigir/ChenZR17}. 
As shown in our experiments, existing deep models fail to beat linear models due to the high-dimensionality of side information and an insufficient number of samples.


To avoid the impact from the high-dimensionality while taking the effectiveness of \ac{VAE}, we propose to learn feature representations from side information.
In this way, the dimensions of the side information correspond to the number of samples rather than the input dimension of deep network. 
To instantiate this idea, in this paper, we propose \ac{cVAE}, which learns to recover user ratings and side information simultaneously through \ac{VAE}.
While user ratings and side information are different sources of information, both are information associated with items. 
Thus, we take ratings from each user and each dimension of side information over all items as the input for \ac{VAE}, so that samples from both sources of information have the same dimensionality (number of items). 
We can then feed ratings and side information into the same inference network and generation network. 
\ac{cVAE} complements the sparse ratings with side information, as feeding side information into the same \ac{VAE} increases the number of samples for training.
The high-dimensionality of side information is not a problem for \ac{cVAE}, as it increases the sample size rather than the network scale.
To account for the heterogeneity of user rating and side information, the final layer of the generation network follows different distributions depending on the type of information.
Training a \ac{VAE} by feeding it side information as input acts like a pre-training step, which is a crucial step for developing a robust deep network. 
Our experiments show that the proposed model,\ac{cVAE}, achieves state-of-the-art performance for top-$N$ recommendation with side information.

\begin{figure}
\begin{tikzpicture}

\tikzstyle{var} = [draw, circle, minimum size=17]
\tikzstyle{obs} = [var,fill=black!30]
    

\begin{scope}[xscale=0.35, yscale=0.35]

\foreach \x/\y in {0/0,0/10,20/0,20/10}{
  \draw (\x-1,\y-1) -- (\x+1,\y-1) -- (\x+1,\y+7) -- (\x-1,\y+7) -- cycle;
  \foreach \i in {0,2,4,6}
    \node at (\x,\i+\y) [obs] {};
}  

\foreach \x/\y in {4/5,16/5}{
  \draw (\x-1,\y-1) -- (\x+1,\y-1) -- (\x+1,\y+7) -- (\x-1,\y+7) -- cycle;
  \foreach \i in {0,2,4,6}
    \node at (\x,\i+\y) [var] {};
}

\foreach \x/\y in {8/10,8/4,12/7}{
  \draw (\x-1,\y-1) -- (\x+1,\y-1) -- (\x+1,\y+3) -- (\x-1,\y+3) -- cycle;
  \foreach \i in {0,2}
    \node at (\x,\i+\y) [var] {};
}

\draw (1,13)[-angle 60] -- (3,9);
\draw (1,3)[-angle 60] -- (3,7);
\draw (5,9)[-angle 60] -- (7,11);
\draw (5,7)[-angle 60] -- (7,5);
\draw (9,11)[-angle 60] -- (11,8.5);
\draw (9,5)[-angle 60] -- (11,7.5);
\draw (13,8)[-angle 60] -- (15,8);
\draw (17,9)[-angle 60] -- (19,13);
\draw (17,7)[-angle 60] -- (19,3);

\draw [red, dashed, very thick] (-2,-1.7) -- (14,-1.7) -- (14,18.3) -- (-2,18.3) -- cycle;
\draw [blue, dashed, very thick] (10,-2.3) -- (22,-2.3) -- (22,17.7) -- (10,17.7) -- cycle;

\node at (6,17) {Inference network};
\node at (14.5,16.5) {Generation network};
\node at (1.5,14) {$\vec{x}$};
\node at (1.5,2) {$\vec{y}$};
\node at (4,3.5) {$\vec{h}_{inf}$};
\node at (8,8.5) {$\vec{\mu}$};
\node at (8,2.5) {$\vec{\sigma}$};
\node at (12,5.5) {$\vec{u},\vec{z}$};
\node at (16,3.5) {$\vec{h}_{gen}$};
\node at (20,8.5) {$\vec{x}$};
\node at (20,-1.5) {$\vec{y}$};

\end{scope}
\end{tikzpicture}
\caption{Collective Variational Autoencoder}
\label{fig:bvae}
\end{figure}
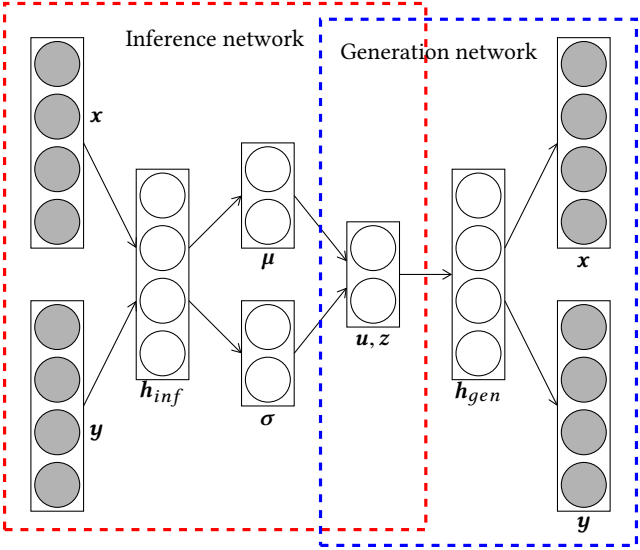

The remainder of the paper is organized as follows. 
We present preliminaries in Section~\ref{sec:prelim}. 
We introduce the \ac{cVAE}model and optimization in Section~\ref{sec:model}. 
Section~\ref{sec:exp} describes the experimental setup and results. 
We review related work in Section~\ref{sec:relat} and conclude in Section~\ref{sec:conclusion}.


\section{Preliminaries} 
\label{sec:prelim}

\subsection{Notation}
\label{sec:notat}

We introduce relevant notation in this section. 
We use $m$, $n$ and $d$ to denote the number of users, items and the dimension of side information, respectively.
We study the problem of top-$N$ recommendation with high-dimensional side information, where $d\gg n$.
We write $X\in\mathbb{R}^{n\times d}$ for the matrix for side information and $Y\in\mathbb{R}^{m\times n}$ for user ratings. 
We summarize our notation in Table~\ref{table:notation}.

\begin{table}[h]
\caption{Notation used in the paper.}
\label{table:notation}
\begin{tabular}{llc}
\toprule
Notation & Description \\
\midrule
$m$ & number of users \\
$n$ & number of items\\
$d$ & dimension of side information \\
$k$ & dimension of latent item representation \\
$N$ & number of recommended items \\
\hline
$X\in\mathbb{R}^{n\times d}$ & matrix of side information \\
$Y\in\mathbb{R}^{n\times m}$ & matrix of user rating \\
$U\in\mathbb{R}^{m\times k}$ & matrix of latent user representation \\
$V\in\mathbb{R}^{n\times k}$ & matrix of latent item representation \\
$Z\in\mathbb{R}^{d\times k}$ & matrix of latent feature representation \\
$\vec{h}_{inf}$ & hidden layer of inference network \\
$\vec{h}_{gen}$ & hidden layer of generation network \\
$\vec{\mu}\in\mathbb{R}^k$ & the mean of latent input representation \\
$\vec{\sigma}\in\mathbb{R}^k$ & the variance of latent user or feature  representation \\
\hline
$f_\phi(\cdot)$ & non-linear transformation of inference network \\
$f_\theta(\cdot)$ & non-linear transformation of generation network \\
$\mu(\cdot)$ & the activation function to get $\vec{\mu}$ \\
$\sigma(\cdot)$ & the activation function to get $\vec{\sigma}$ \\
$\varsigma(\cdot)$ & the sigmoid function \\
\bottomrule
\end{tabular}
\end{table}

\subsection{Linear models for top-$N$ recommendation}

\ac{SLIM}~\cite{DBLP:conf/icdm/NingK11} achieves state-of-the-art performance for top-$N$ recommendation. 
\ac{SLIM} learns to reproduce the user rating marix $Y$ through:
\begin{displaymath}
	Y\sim YW.
\end{displaymath}
Here, $W\in\mathbb{R}^{n\times n}$ is the coefficient matrix, which is analogous to the item similarity matrix.
The performance of \ac{SLIM} is heavily affected by the rating sparsity~\cite{DBLP:conf/kdd/KabburNK13}.
Side information ha been utilized to overcome this issue~\cite{DBLP:conf/recsys/NingK12,DBLP:conf/ijcai/ZhaoXG16,DBLP:conf/sigir/ChenZR17}. 
As a typical example of a method that uses side information, \ac{cSLIM} learns $W$ from both user rating and side information. Specifically, $X,Y$ are both reproduced through:
\begin{displaymath}
Y\sim YW,\quad X\sim XW.
\end{displaymath}
\ac{cSLIM} learns the coefficient matrix $W$ collectively from both side information $X$ and user rating $Y$, a strategy that can help to overcome rating sparsity by side information. 
However, \ac{cSLIM} is restricted by the fact that it is a linear model, which has limited model capacity. 

\subsection{Autoencoders for collaborative filtering}

Recently, autoencoders have been used to address \ac{CF} problems~\cite{DBLP:conf/www/SedhainMSX15,DBLP:conf/recsys/StrubGM16,DBLP:conf/wsdm/WuDZE16,DBLP:journals/nn/ZhuangZQSXH17}. 
Autoencoders are neural networks popularized by~\citet{kramer1991nonlinear}. 
They are unsupervised networks where the output of the network aims to be a reconstruction of  the input.

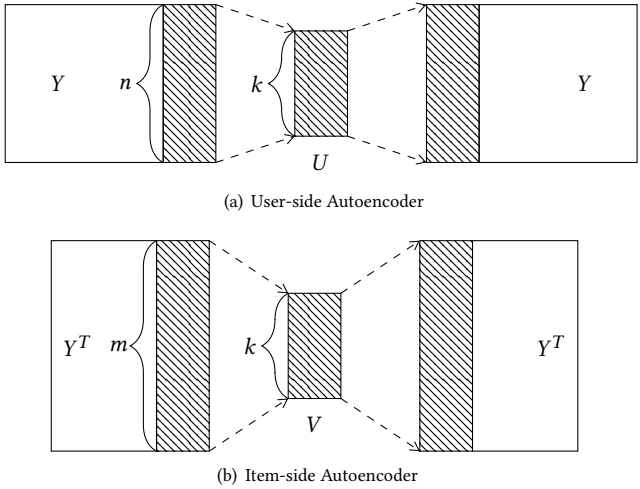
\begin{figure}
\centering
\subfigure[User-side Autoencoder]{
\begin{tikzpicture}[x=0.7cm, y=0.7cm]

\draw (0,3) rectangle (3,0);
\draw (9,3) rectangle (12,0);
\draw[pattern=north west lines] (3,3) rectangle (4,0);
\draw[pattern=north west lines] (8,3) rectangle (9,0);
\draw[pattern=north west lines] (5.5,2.5) rectangle (6.5,0.5);
\draw[-angle 60,dashed] (4,3) -- (5.5,2.5);
\draw[-angle 60,dashed] (4,0) -- (5.5,0.5);
\draw[-angle 60,dashed] (6.5,2.5) -- (8,3);
\draw[-angle 60,dashed] (6.5,0.5) -- (8,0);

\node at (1,1.5) {$Y$};
\node at (11,1.5) {$Y$};
\node at (6,0) {$U$};

\draw[decorate,decoration={brace,amplitude=10pt}]
(5.5,.5) -- (5.5,2.5) node [midway,xshift=-.5cm] {$k$};

\draw[decorate,decoration={brace,amplitude=10pt}]
(3,0) -- (3,3) node [midway,xshift=-.5cm] {$n$};
\end{tikzpicture}
\label{subfig:uae}
}
\subfigure[Item-side Autoencoder]{
\begin{tikzpicture}[x=0.7cm, y=0.7cm]
\usetikzlibrary{arrows}
\usetikzlibrary{patterns}

\draw (1,4) rectangle (3,0);
\draw (9,4) rectangle (11,0);
\draw[pattern=north west lines] (3,4) rectangle (4,0);
\draw[pattern=north west lines] (8,4) rectangle (9,0);
\draw[pattern=north west lines] (5.5,3) rectangle (6.5,1);
\draw[-angle 60,dashed] (4,4) -- (5.5,3);
\draw[-angle 60,dashed] (4,0) -- (5.5,1);
\draw[-angle 60,dashed] (6.5,3) -- (8,4);
\draw[-angle 60,dashed] (6.5,1) -- (8,0);

\node at (1.5,2) {$Y^T$};
\node at (10.5,2) {$Y^T$};
\node at (6,0.5) {$V$}; 

\draw[decorate,decoration={brace,amplitude=10pt}]
(5.5,1) -- (5.5,3) node [midway,xshift=-.5cm] {$k$};

\draw[decorate,decoration={brace,amplitude=10pt}]
(3,0) -- (3,4) node [midway,xshift=-.5cm] {$m$};

\end{tikzpicture}
\label{subfig:iae}
}
\caption{Autoencoders for collaborative filtering}
\end{figure}

In the context of \ac{CF}, the autoencoder is fed with incomplete rows (resp.\ columns) of the user rating matrix $Y$. 
It then outputs a vector that predicts the missing entries. 
These approaches perform a non-linear low-rank approximation of $Y$ in two different ways,  using a \ac{UAE} (Figure~\ref{subfig:uae}) or \ac{IAE} (Figure~\ref{subfig:iae}), which recover $Y$ respectively through:
\begin{displaymath}
	U\sim f(Y),\quad Y\sim g(U),
\end{displaymath}
and
\begin{displaymath}
	V\sim f(Y^T),\quad Y^T\sim g(V),
\end{displaymath}
where $U\in\mathbb{R}^{m\times k}$ is the user representation and $V\in\mathbb{R}^{n\times k}$ is the item representation. 
Moreover, $f(\cdot)$ and $g(\cdot)$ are the encode network and decode network, respectively. 
\acp{UAE} encode $Y$ to learn a user latent representation $U$ and then recover $Y$ from $U$. 
In contrast, \acp{IAE} encode the transpose of $Y$ to learn item latent representation $V$ and then recover the transpose of $Y$ from $V$.
Note that \acp{UAE} work in a similar way as \ac{SLIM}, as both can be viewed as reproducing $Y$ through $Y\sim g(f(Y))$, which also captures item similarities.

When side information associated with items is available, the \ac{FAE} is utilized to learn item representations:
\begin{displaymath}
V'\sim f(X),\quad X\sim g(V'),
\end{displaymath}
where $V'\in\mathbb{R}^{n\times k}$ is the item representation.
Existing hybrid methods incorporate \ac{FAE} with \ac{IAE} as both learn item representations. 
However, this way of incorporating side information needs to estimate two separate \acp{VAE}, which is not an effective way to address rating sparsity. 
They are also vulnerable to the high dimensionality of side information.


\section{Method} \label{sec:model}

In this section, we propose a new way to incorporate side information with user ratings by combining the effectiveness of both \ac{cSLIM} and autoencoders.
We propose to reproduce $X$ by a \ac{FAE} and $Y$ by a \ac{UAE}. 
In this way, the input for autoencoders of both $X$ and $Y$ are of the same dimension, i.e., the number of items $n$. 
Thus, we can feed $X$ and $Y$ into the same autoencoder rather than two different autoencoders, which helps to overcome rating sparsity. 

\subsection{Collective variational autoencoder}

We propose a \acf{cVAE} to generalize the linear models for top-$N$ recommendation with side information to non-linear models, by taking advantage of \acfp{VAE}.
Specifically, we propose to recover $X,Y$ through
\begin{displaymath}
\begin{split}
& U\sim f_\phi(Y),\quad Y\sim f_\theta(U), \\
& Z\sim f_\phi(X),\quad X\sim f_\theta(Z),
\end{split}
\end{displaymath}
where $f_\phi(\cdot)$ and $f_\theta(\cdot)$ correspond to the inference network and generation network parameterized by $\phi$ and $\theta$, respectively. 
An overview of \ac{cVAE} is depicted in Figure~\ref{fig:bvae}.
Unlike previous work utilizing \acp{VAE}, the proposed model encodes and decodes user rating and side information through the same inference and generation networks. 
Our model can be viewed as a non-linear generalization of \ac{cSLIM}, so as to learn  item similarities collectively from user ratings and side information. 
While user ratings and side information are two different types of information, \ac{cSLIM} fails to distinguish them. 
In contrast, \ac{cVAE} assumes the output of the generation network to follow different distributions according to the type of input it has been fed.

Next, we describe the \ac{cVAE} model in detail. 
Following common practice for \ac{VAE}, we first assume the latent variables $\vec{u}$ and $\vec{z}$ to follow a Gaussian distribution:
\begin{displaymath}
	\vec{u}\sim \mathcal{N}(0,I),\quad\vec{z}\sim \mathcal{N}(0,I),
\end{displaymath}
where $I\in\mathbb{R}^{k\times k}$ is an identity matrix.
While $X$ and $Y$ are fed into the same network, we would like to distinguish them via different distributions. 
In this paper, we assume that $Y$ is binarized to capture implicit feedback, which is a common setting for top-$N$ recommendation~\cite{DBLP:conf/icdm/NingK11}. 
Thus we follow~\citet{DBLP:conf/cikm/LeeSM17} and assume that the rating of user $j$ over all items follows a Bernoulli distribution:
\begin{displaymath}
	\vec{y}_j\mid \vec{u}_j\sim Bernoulli(\varsigma(f_\theta(\vec{u}_j))),
\end{displaymath}
where $\varsigma(\cdot)$ is the sigmoid function.
This defines the loss function when feeding user rating as input, i.e., the logistic log-likelihood for user $j$:
\begin{equation}\label{eq:bce}
\log p_\theta(\vec{y}_j\mid \vec{u}_j)=\sum_{i=1}^n y_{ji}\log \varsigma(f_{ji})+(1-y_{ji})\log\rbr{1-\varsigma(f_{ji})},
\end{equation}
where $f_{ji}$ is the $i$-th element of the vector $f_\theta(\vec{u}_j)$ and $f_\theta(\vec{u}_j)$ is normalized through a sigmoid function so that $f_{ji}$ is within $(0,1)$. 

For side information, we study numerical features so that we assume the $j$-th dimension of side information from all items follows a Gaussian distribution:
\begin{displaymath}
	\vec{x}_j\mid \vec{z}_j\sim \mathcal{N}(f_\theta(\vec{z}_j), I).
\end{displaymath} 
This defines the loss function when feeding side information as input, i.e., the Gaussian log-likelihood for dimension $j$:
\begin{equation}\label{eq:rmse}
\log p_\theta(\vec{x}_j\mid \vec{z}_j)=\sum_{i=1}^n -\frac{1}{2}(x_{ji}-f_{ji})^2,
\end{equation}
where $f_{ji}$ is the $i$-th element of vector $f_\theta(\vec{z}_j)$. Note that although we assume $\vec{x}$ and $\vec{y}$ to be generated from $\vec{z}$ and $\vec{u}$ respectively, the generation has shared parameters $\theta$.

The generation procedure is summarized as follows:
\begin{enumerate}
	\item for each user $j=1,\ldots,m$:
	\begin{enumerate}
	\item draw $\vec{u}_j\sim\mathcal{N}(0,I)$;
	\item draw $\vec{y}_j\sim Bernoulli\rbr{\varsigma(f_\theta(\vec{u}_j))}$
	\end{enumerate}
	\item for each dimension of side information $j=1,\ldots,d$:
	\begin{enumerate}
	  \item draw $\vec{z}_j\sim\mathcal{N}(0,I)$;
	  \item draw $\vec{x}_j\sim\mathcal{N}(f_\theta(\vec{z}_u),I)$.
	\end{enumerate}
\end{enumerate}

\noindent%
Once the \ac{cVAE} is trained, we can generate recommendations for each user $j$ with items ranked in descending order of $f_\theta(\vec{u}_j)$. 
Here, $\vec{u}_j$ is calculated as $\vec{u}_j=\mu(f_\phi(\vec{y}_j))$, that is, we take the mean of $\vec{u}_j$ for prediction.

Next, we discuss how to perform inference for \ac{cVAE}.

\subsection{Variational inference}

The log-likelihood of \ac{cVAE} is intractable due to the non-linear transformations of the generation network. 
Thus, we resort to variational inference to approximate the distribution. 
Variational inference approximates the true intractable posterior with a simpler variational distribution $q(U,Z)$.
We follow the mean-field assumption~\cite{xing2002generalized} by setting $q(U,Z)$ to be a fully factorized Gaussian distribution:
\begin{displaymath}
\begin{split}
q(U,Z)={}&\prod_{j=1}^{m}  q(\vec{u}_j)\prod_{j=1}^d q(\vec{z}_j),\\
q(\vec{u}_j)={}&\mathcal{N}(\vec{\mu}_j,\text{diag}(\vec{\sigma}^2_j)),\\
q(\vec{z}_j)={}&\mathcal{N}(\vec{\mu}_{m+j},\text{diag}(\vec{\sigma}^2_{m+j})),
\end{split}
\end{displaymath}
While we can optimize $\{\vec{\mu}_j,\vec{\sigma}_j\}$ by minimizing the Kullback-Leiber divergence $\mathbb{KL}(q_\phi\parallel p_\theta)$, the number of parameters to learn grows with the number of users and dimensions of side information. 
This can become a bottleneck for real-world recommender systems with millions of users and high-dimensional side information. 
The \ac{VAE} replaces individual variational parameters with a data-dependent function through an inference network parameterized by $\phi$, i.e., $f_\phi$, where $\vec{\mu}_j$ and $\vec{\sigma}_j$ are generated as:
\begin{displaymath}
\begin{split}
& \vec{\mu}_j=\mu(f_\phi(\vec{u}_j)),\quad \vec{\sigma}_j=\sigma(f_\phi(\vec{p}_j)), \quad\forall j=1,\ldots,m \\
& \vec{\mu}_{m+j}=\mu(f_\phi(\vec{z}_j)),\quad \vec{\sigma}_{m+j}=\sigma(f_\phi(\vec{z}_j)),\quad\forall j=1,\ldots,d.
\end{split}
\end{displaymath}
Putting together $p_\phi(\vec{z}\mid\vec{x})$ and $p_\phi(\vec{u}\mid\vec{y})$ with $p_\theta(\vec{x}\mid\vec{z})$ and $p_\theta(\vec{y}\mid\vec{u})$ forms the proposed \ac{cVAE} (Figure~\ref{fig:bvae}).

We follow to derive the \ac{ELBO}:
\begin{equation}\label{eq:elbo}
\mathcal{L}(q)=\mathbb{E}_{q_\phi}\sbr{\log p_\theta(X,Y\mid U,Z)}-\mathbb{KL}\rbr{q_\phi\parallel p(U,Z)}.
\end{equation}
We use a Monte Carlo gradient estimator~\cite{DBLP:conf/icml/PaisleyBJ12} to infer the expectation in Equation~\eqref{eq:elbo}. We draw $L$ samples of $\vec{u}_j$ and $\vec{z}_j$ from $q_\phi$ and perform stochastic gradient ascent to optimize the \ac{ELBO}. In order to take gradients with respect to $\phi$ through sampling, we follow the reparameterization trick~\cite{DBLP:journals/corr/KingmaW13} to sample $\vec{u}_j$ and $\vec{z}_j$ as:
\begin{displaymath}
\begin{split}
& \vec{u}_j^{(l)}=\mu(f_\phi(\vec{y}_j))+\vec{\epsilon}_1^{(l)}\odot\sigma(f_\phi(\vec{y}_j)), \\
& \vec{z}_j^{(l)}=\mu(f_\phi(\vec{x}_j))+\vec{\epsilon}_2^{(l)}\odot\sigma(f_\phi(\vec{x}_j)), \\
& \vec{\epsilon}_1^{(l)}\sim\mathcal{N}(0,I),\quad \vec{\epsilon}_2^{(l)}\sim\mathcal{N}(0,I) .
\end{split}
\end{displaymath}
As the $\mathbb{KL}$-divergence can be analytically derived~\cite{DBLP:journals/corr/KingmaW13}, we can then rewrite $\mathcal{L}(q)$ as:
\begin{equation}\label{eq:objective}
\begin{split}
\mathcal{L}(q)={}&\frac{1}{L}\sum_{l=1}^L\rbr{\sum_{j=1}^m\log p_\theta(\vec{y}_j\mid\vec{u}_j^{(l)})+\sum_{j=1}^d\log p_\theta(\vec{x}_j\mid\vec{z}_j^{(l)})}+\\
& \phantom{\left(\frac{1}{L}\sum_{l=1}^L\right.}\sum_{j=1}^{d+m}\rbr{1+2\log(\vec{\sigma}_j)-\vec{\mu}_j^2-\vec{\sigma}_j^2}.
\end{split}
\end{equation} 
We then maximize \ac{ELBO} given in Equation~\eqref{eq:objective} to learn $\theta$ and $\phi$.

\subsection{Implementation details}

We discuss the implementation of \ac{cVAE} in detail. 
As  we feed the user rating matrix $Y$ and the item side information $X$ through the same input layer with $n$ neurons, we need to ensure that the input from both types of information are of the same format. 
In this paper, we assume that user ratings are binarized to capture implicit feedback and that side information is represented as a bag-of-words.
We propose to train \acp{cVAE} through a two-phase algorithm. 
We first feed it side information to train, which works as pre-training. 
We then refine the \ac{VAE} by feeding user ratings.
We follow the typical setting by taking $f_\theta$ as a \ac{MLP}; $f_\phi$ is also taken to be a \ac{MLP} of the identical network structure with $f_\theta$. 
We also introduce two parameters, i.e., $\alpha$ and $\beta$, to extend the model and make it more suitable for the recommendation task.

\subsubsection{Parameter $\alpha$}
For both $X$ and $Y$, most entires are zeros. 
We introduce a parameter $\alpha$ to balance between positive samples and negative samples. 
Specifically, the loss functions in Equation~\eqref{eq:bce} and \eqref{eq:rmse} become
\begin{displaymath}
\log p_\theta(\vec{y}_j\mid \vec{u}_j)\simeq \alpha\sum_{i\in\mathcal{R}_u^+}\log \varsigma(f_{ji})+\sum_{i\in\mathcal{R}_u^-}\log\rbr{1-\varsigma(f_{ji})}
\end{displaymath}
and
\begin{displaymath}
\log p_\theta(\vec{x}_j\mid \vec{z}_j)\simeq-\frac{\alpha}{2}\sum_{i\in\mathcal{R}_u^+} (1-f_{ji})^2-\frac{1}{2}\sum_{i\in\mathcal{R}_u^-} f_{ji}^2.
\end{displaymath}

\subsubsection{Parameter $\beta$}
We can adopt different perspectives about the \ac{ELBO} derived in Equation~\eqref{eq:elbo} as: the first term can be interpreted as the reconstruction error, while the second $\mathbb{KL}$ term can be viewed as regularization. 
The \ac{ELBO} is often over-regularized for recommendation tasks~\cite{DBLP:conf/www/LiangKHJ18}. 
Therefore, a parameter $\beta$ is introduced to control the strength of regularization, so that the \ac{ELBO} becomes:
\begin{displaymath}
\mathcal{L}(q)=\mathbb{E}_{q_\phi}\sbr{\log p_\theta(X,Y\mid U,Z)}-\beta\mathbb{KL}\rbr{q_\phi\parallel p(U,Z)}.
\end{displaymath}
We propose to train the \ac{cVAE} in two phases. 
We first pre-train the \ac{cVAE} by feeding it side information only. 
We then refine the model by feeding it user ratings. 
While \citet{DBLP:conf/www/LiangKHJ18} suggests to set $\beta$ small to avoid over-regularization, we opt for a larger value for $\beta$ during refinement, for two reasons:
\begin{inparaenum}
\item the model is effectively pre-trained with side information; it would be reasonable to require the posterior to comply more with this prior; and
\item refinement with user ratings can easily overfit due to the sparsity of ratings; it would be reasonable to regularize heavier so as to avoid overfitting.
\end{inparaenum}


\section{Experiments}
\label{sec:exp}

\subsection{Experimental setup}
\subsubsection{Dataset}
We conduct experiments on two datasets, \game and \sport, constructed from different categories of Amazon products~\cite{DBLP:conf/recsys/McAuleyL13}. 
For each category, the original dataset contains transactions between users and items, indicating implicit user feedback. 
The statistics of the datasets are presented in Table~\ref{tab:dataset}. 
We use the product reviews as item featured. 
We extract unigram features from the review articles and remove stopwords. 
We represent each product item as a bag-of-words feature vector.

\begin{table}[h]
    \centering
    \caption{Statistics of the datasets used.}\label{tab:dataset}
	\begin{tabular}{lrrrrr}
	\toprule
    \textbf{Dataset} & \#User & \#Item & \#Rating & \#Dimension & \#Feature \\
	\hline
	\game  & \numprint{5195} & \numprint{7163} & \numprint{96316} & \numprint{20609} & \numprint{5151174} \\
	\sport & \numprint{5653} & \numprint{11944} & \numprint{86149} & \numprint{31282} & \numprint{3631243} \\
	\toprule
	\end{tabular}
\end{table}

\subsubsection{Methods for comparison}
We contrast the performance of \cvae with that of existing existing \ac{VAE}-based methods for \ac{CF}: \cfvae~\cite{DBLP:conf/kdd/LiS17} and \rvae~\cite{DBLP:conf/www/LiangKHJ18}. 
Note that the performance of \cfvae will be affected greatly by the high-dimensionality of side information. 
Besides, as \cfvae is designed originally for the rating prediction task, the recommendations provided by \cfvae will be less effective. 
While \rvae is effective for top-$N$ recommendation, it suffers from rating sparsity as side information is not utilized. 
 
We also compare with the state-of-the-art linear model for top-$N$ recommendation with side information, i.e., \cslim~\cite{DBLP:conf/recsys/NingK12}.
By comparing with \cslim, we can evaluate the capacity of \cvae as it can be regarded as a deep extension of \cslim.
We also compare with \fvae, which is the pre-trained model of \cvae with side information only.
Note that \cvae is the refinement over \fvae by user rating.

For all the \ac{VAE}-based methods, we follow~\citet{DBLP:journals/corr/KingmaW13} to set the batch size as 100 so that we can set $L=1$.
We choose a two-layer network architecture for the inference network and generation network. 
For \cfvae and \rvae, the scale is 200-100 for inference network and 100-200 for generation network. 
For \fvae and \cvae, the scale is 1000-100 and 100-1000, respectively. The reason that the network scale for \cfvae and \rvae is relatively smaller is that 
\begin{inparaenum}
\item the input for \cfvae is high-dimensional with relatively fewer samples; and
\item the input for \rvae is sparse, which easily overfits for larger network scale.
\end{inparaenum}
In comparison, we can select more hidden neurons for \fvae as it takes each dimension of the features over all items as input, so that the input for the network has relatively fewer dimensions and the number of samples is sufficient.
This is similar with \cvae, which uses side information to overcome rating sparsity.

\subsubsection{Evaluation method}
To evaluate the performance of top-$N$ recommendation, we split the user rating matrix $R$ into $R_\mathit{train},R_\mathit{valid}$ and $R_\mathit{test}$, respectively, for training the model, selecting parameters and testing the recommendation accuracy.
Specifically, for each user, we randomly hold 10\% of the ratings in the validation set and 10\% in the test set and put the other ratings in the training set.
For each user, the unrated items are sorted in decreasing order according to the predicted score and the first $N$ items are returned as the top-$N$ recommendations for that user.  

Given the list of top-$N$ recommended items for user $u$, \ac{Pre@N} and \ac{Rec@N} are defined as
\begin{displaymath}
\begin{split}
Rre@N &=\frac{\size{\text{relevant items}\cap \text{recommended items}}}{N}, \\
Rec@N &=\frac{\size{\text{relevant items}\cap \text{recommended items}}}{\size{\text{relevant items}}}.
\end{split}
\end{displaymath}
\ac{AP@N} is a ranked precision metric that gives larger credit to correctly recommended items in the top-$N$ ranks. \ac{AP@N} is defined as the average of precisions computed at all positions with an adopted item, namely
\begin{displaymath}
	AP@N=\frac{\sum_{k=1}^N Pre@k\times rel(k)}{\min\set{N,\size{\text{relevant items}}}},
\end{displaymath}
where Pre@k is the precision at cut-off $k$ in the top-$N$ recommended list. 
Here, $rel(k)$ is an indicator function
\begin{displaymath}
rel(k)=
\begin{cases}
1 & \text{if the item ranked at $k$ is relevant}, \\
0 & \text{otherwise}.	
\end{cases}
\end{displaymath}
\ac{MAP@N} is defined as the mean of the AP scores for all users.
Following~\citet{DBLP:conf/wsdm/WuDZE16}, the list of recommended items is evaluated with $R_\mathit{test}$ using \ac{Rec@N} and \ac{MAP@N}.

\subsection{Experimental results}

\subsubsection{Parameter selection}
To compare the performance of alternative top-$N$ recommendation methods, we first select parameters for all the methods through validation. 
Specifically, for \cslim, we select $\alpha,\beta$ and $\lambda$ from $0$, $10^{-4}$, $10^{-3}$, $10^{-2}$, $10^{-1}$, $1$, $10$.
For \cfvae, we select $\lambda_u$ from $0,10^{-4},10^{-3},10^{-2},10^{-1},1,10$ and $\lambda_v$ from $0,0.1,0.2,\ldots,1$.
For \rvae and \fvae, we select $\alpha$ from $1,2,\ldots,10$ and $\beta$ from $0,0.1,0.2,\ldots,1$. For \cvae, we select $\alpha$ from $1,2,\ldots,10$ and $\beta$ from $0,0.5,1,\ldots,3$.
Note that we tune $\beta$ with larger values to possibly regularize heavier during the refinement. 

The result of parameter selection is shown in Table~\ref{tab:parameter}.

\begin{table}[h]
\centering
\caption{Parameter selection.}
\label{tab:parameter}
\begin{tabular}{lll}
\toprule
\multirow{2}{*}{\textbf{Method}} & \multicolumn{2}{c}{\textbf{Parameters}} \\
\cmidrule{2-3}
& \game & \sport \\
\midrule
\cslim & $\alpha=10^{-2},\beta=10^{-3},\lambda=1$ & $\alpha=0,\beta=10,\lambda=10$ \\
\cfvae & $\lambda_u=10,\lambda_v=0.8$ & $\lambda_u=10,\lambda_v=1$ \\
\rvae & $\alpha=4,\beta=0.1$ & $\alpha=8,\beta=0.1$ \\
\fvae & $\alpha=6,\beta=0.1$ & $\alpha=5,\beta=0.9$ \\
\cvae & $\alpha=2,\beta=2$ & $\alpha=1,\beta=2$ \\
\bottomrule
\end{tabular}	
\end{table}
 
\subsubsection{Performance comparison}
We present the results in terms of \ac{Rec@N} and \ac{MAP@N} in Table~\ref{tab:result}, where $N$ is respectively set as $N=5,10,15,20$.
We show the best score in boldface. 
We attach asterisks to the best score if the improvement over the second best score is statistically significant; to this end, we conducted two-sided tests for the null hypothesis that \ac{cVAE} and the second best have identical average values; we use one asterisk if $p < 0.1$ and two asterisks if $p<0.05$.

\newcommand{\StatSig}[1]{\smash{\rlap{#1}}}

\begin{table*}
\centering
\caption{Comparison of top-$N$ recommendation performance. * indicates a statistically significant difference between \cvae and the best performing baseline, where * indicates $p<0.1$ and ** indicates $p<0.05$}
\label{tab:result}
\begin{tabular}{lccccccccc}
    \toprule
  & Rec@5 & Rec@10 & Rec@15 & Rec@20 && MAP@5 & MAP@10 & MAP@15 & MAP@20 \\
  \textbf{Method} & \multicolumn{9}{c}{\game} \\
  \cmidrule{2-10}
  \cslim & 0.0761 & 0.1162 & 0.1474 & 0.1734 && 0.0590 & 0.0337 & 0.0240 & 0.0188 \\
  \cfvae & 0.0685 & 0.1065 & 0.1359 & 0.1608 && 0.0519 & 0.0298 & 0.0212 & 0.0165 \\
  \rvae & 0.0137 & 0.0206 & 0.0270 & 0.0375 && 0.0106 & 0.0060 & 0.0043 & 0.0034 \\
  \fvae & 0.0495 & 0.0796 & 0.1072 & 0.1276 && 0.0390 & 0.0230 & 0.0167 & 0.0131 \\
  \cvae & \textbf{0.0858}\StatSig{*} & \textbf{0.1376}\StatSig{**} & \textbf{0.1731}\StatSig{**} & \textbf{0.2081}\StatSig{**} && \textbf{0.0668}\StatSig{*} & \textbf{0.0394}\StatSig{**} & \textbf{0.0279}\StatSig{**} & \textbf{0.0218}\StatSig{**} \\
  \midrule
  & \multicolumn{9}{c}{\sport} \\
  \cmidrule{2-10}
  \cslim & 0.0419 & 0.0622 & 0.0776 & 0.0911 && 0.0263 & 0.0148 & 0.0104 & 0.0080 \\
  \cfvae & 0.0315 & 0.0512 & 0.0639 & 0.0768 && 0.0206 & 0.0119 & 0.0084 & 0.0065 \\
  \rvae & 0.0171 & 0.0249 & 0.0328 & 0.0390 && 0.0109 & 0.0062 & 0.0044 & 0.0034 \\
  \fvae & 0.0284 & 0.0437 & 0.0602 & 0.0732 && 0.0190 & 0.0109 & 0.0078 & 0.0061 \\
  \cvae & \textbf{0.0441} & \textbf{0.0655} & \textbf{0.0857}\StatSig{*} & \textbf{0.1035}\StatSig{*} && \textbf{0.0268} & \textbf{0.0152} & \textbf{0.0107} & \textbf{0.0084} \\
  \bottomrule
\end{tabular}
\end{table*}

As shown in Table~\ref{tab:result}, \cvae outperforms other methods according to all metrics and on both datasets. 
The improvement is also significant in many settings. 
A general trend is revealed that the significance of improvements become more evident when $N$ gets larger.
Note that the other three methods utilizing \ac{VAE} are less effective with high-dimensional side information. 
Actually, they even fail to beat linear models.
In contrast, \cvae improves over \cslim by using \ac{VAE} for non-linear low-rank approximation.
This demonstrates the effectiveness of our proposed \ac{cVAE} model.

Specifically, on the \game dataset, \cvae shows significant improvements over the state-of-the-art methods. 
Apart from \cvae, \cfvae provides the best recommendation among all \ac{VAE}-based \ac{CF} methods, although it fails to beat \cslim. 
This is followed by \fvae, which utilizes side information only. \rvae performs the worst, due to the rating sparsity.

On the \sport dataset, significant improvements can only be observed for Rec@15 and Rec@20. 
The results yield an interesting insight. 
If we look at the parameter selection for \cslim, we can see that $\alpha$ is set to 0, which means \cslim performs the best recommendation when no side information is utilized. 
This does not necessarily mean that the side information of \sport is useless for recommendation. Actually, \fvae provides acceptable recommendations by utilizing side information only. Therefore, the way of incorporating side information by \cslim is not effective. In comparison, \cvae improves over \cslim by utilizing side information.

\subsubsection{Effect of the number of recommended items}
\begin{figure*}
\centering
\subfigure[Rec@N, \game]{
\includegraphics[width=.45\textwidth]{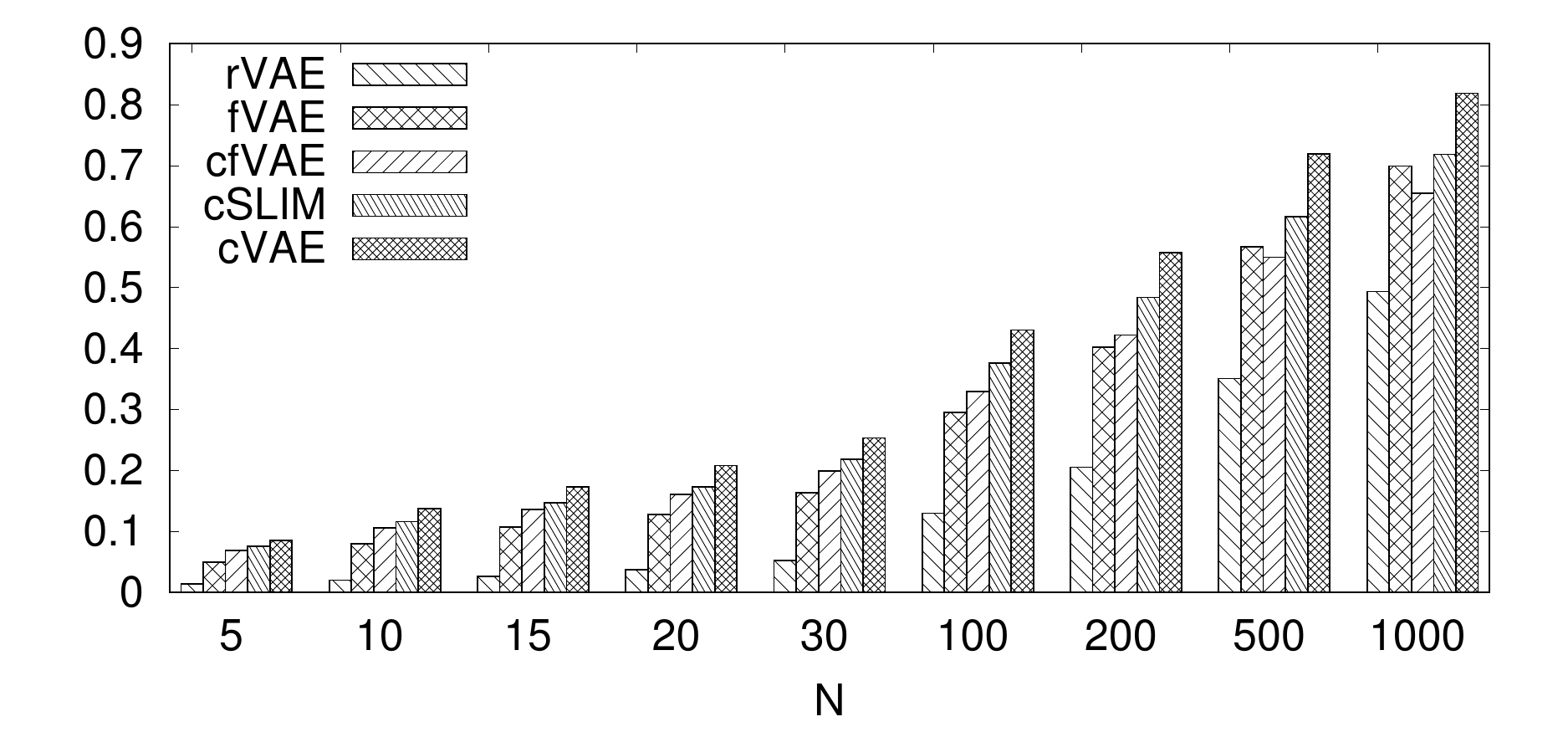}
\label{subfig:recall-game}
}
\goodgap
\subfigure[MAP@N, \game]{
\includegraphics[width=.45\textwidth]{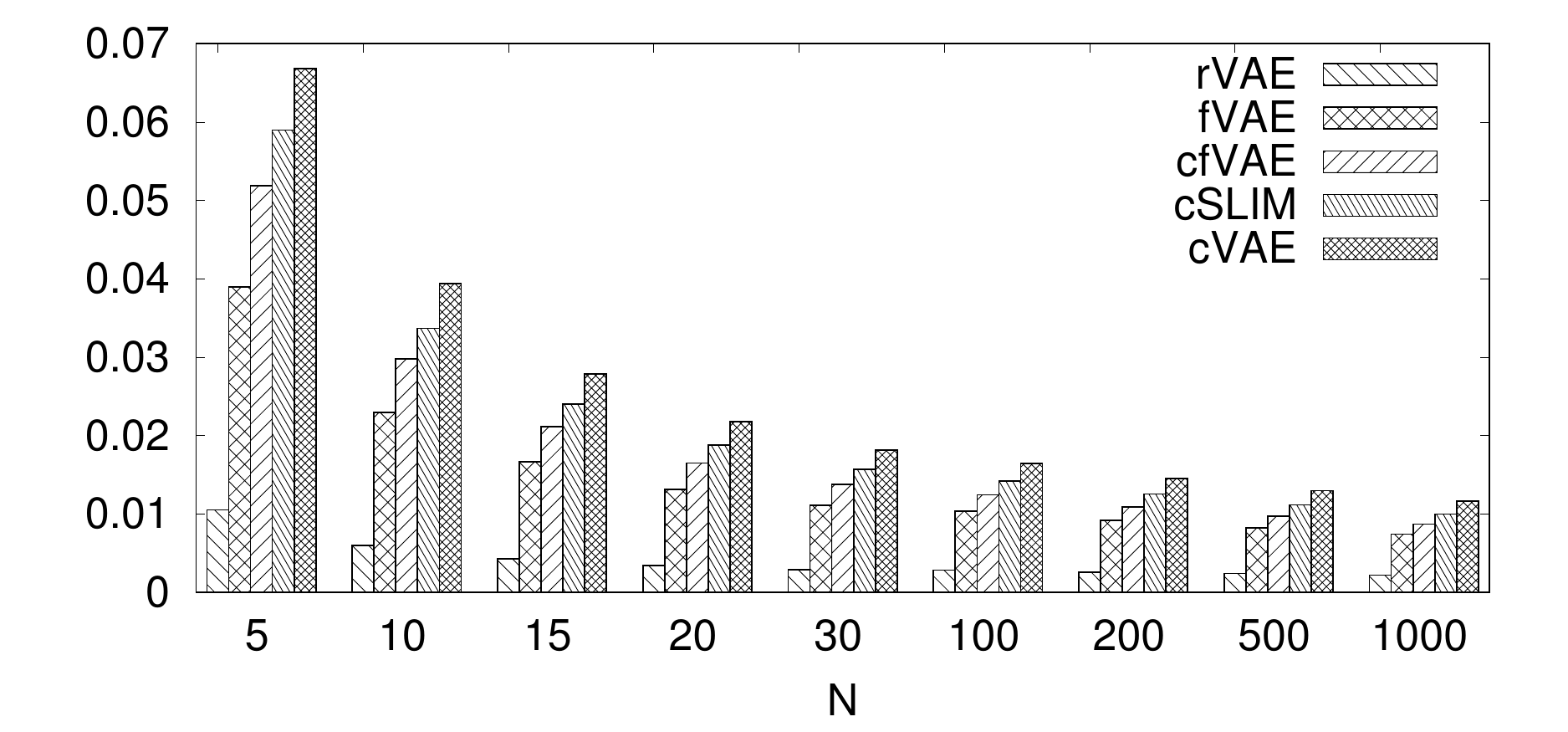}
\label{subfig:map-game}
}
\goodgap
\subfigure[Rec@N, \sport]{
\includegraphics[width=.45\textwidth]{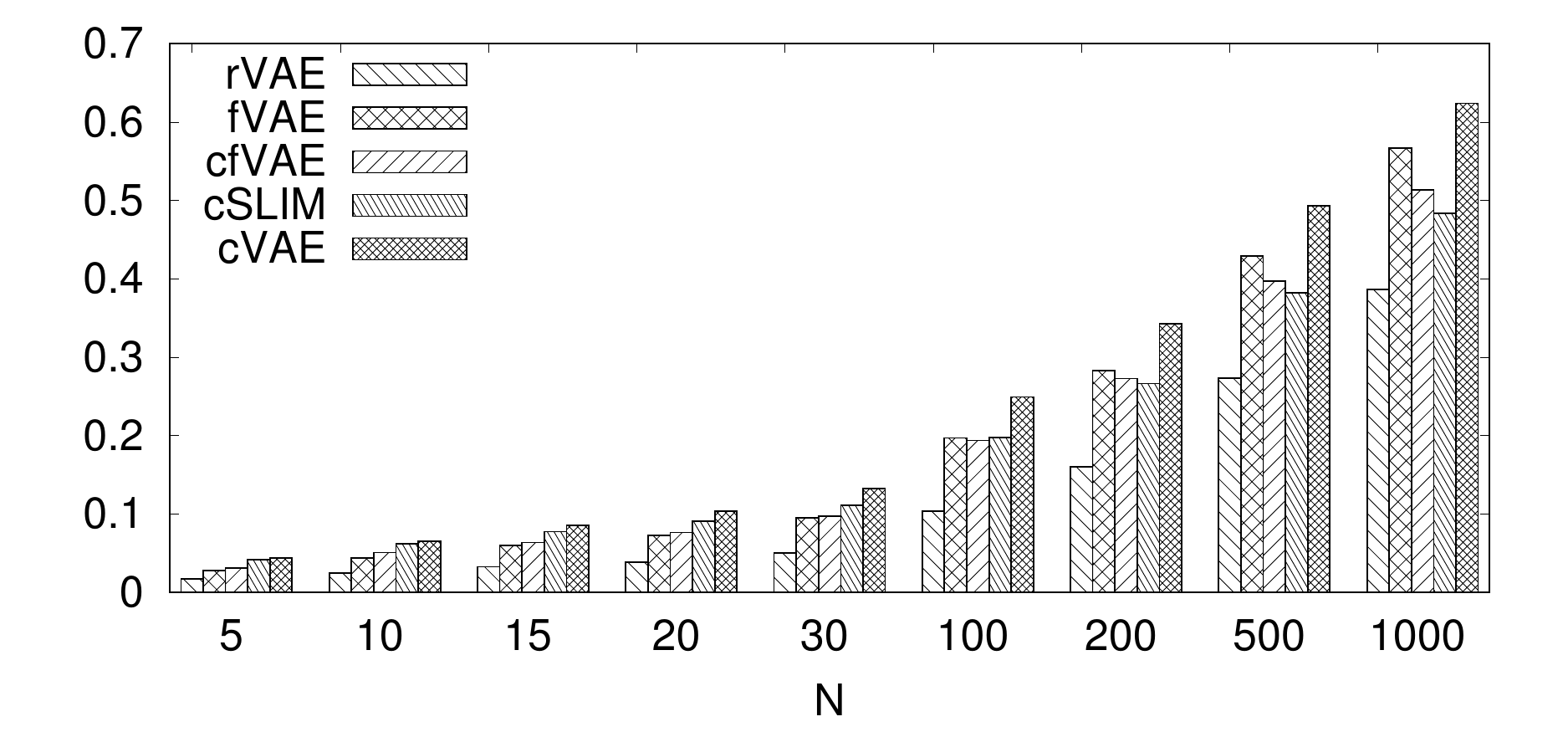}
\label{subfig:recall-sport}
}
\goodgap
\subfigure[MAP@N, \sport]{
\includegraphics[width=.45\textwidth]{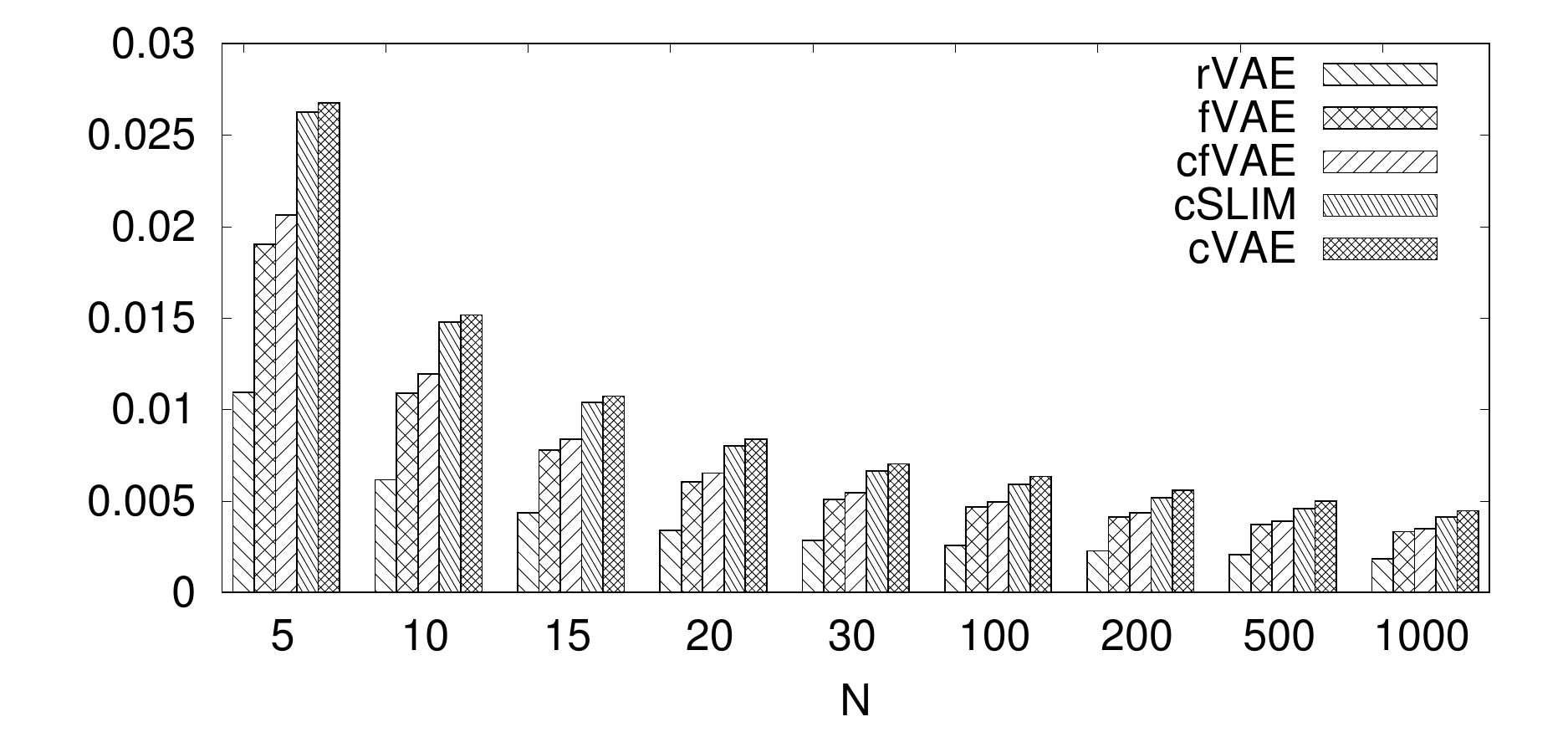}
\label{subfig:map-sport}
}
\caption{Performance of top-$N$ recommendation}
\label{fig:topn}
\end{figure*}

Table~\ref{tab:result} reveals a possible trend that the recommendation improvement becomes more evident when more items are recommended. 
We use Figure~\ref{fig:topn} to illustrate this, where $N$ is increased from 5 to 1000. 

As depicted in Figure~\ref{subfig:recall-game}, the gaps between \cvae and other methods is getting larger with the growth of $N$. It is interesting to note that \fvae surpasses \cfvae when $N=500$ and $N=1000$. This further demonstrates the effectiveness of a pre-train phase with side information proposed in this model.

In Figure~\ref{subfig:recall-sport}, both \fvae and \cfvae outperform \cslim when $N\geq 200$, and \fvae outperforms \cfvae when $N\geq 100$. 
This shows that deep models are superior to linear models when more items are recommended. 
In comparison, the improvement achieved by \cvae is more evident when $N\geq 100$, and the gap between \cvae and the second best method is always substantial.

On the other hand, the performance w.r.t. MAP@N does not show big differences when $N$ grows. Note that on the \game dataset (Figure~\ref{subfig:map-game}), \cvae performs much better than \cslim when $N$ is small. 
The improvement becomes less evident when $N$ grows. 


\section{Related Work}\label{sec:relat}

We review related work on linear models for top-$N$ recommendation with side information and on deep models for collaborative filtering.

\subsection{Top-$N$ recommendation with side information}
Various methods have been developed to incorporate side information in recommender systems. 
Most of these methods have been developed in the context of the rating prediction problem, whereas the top-$N$ recommendation problem has received less attention. 
In the rest of this section we only review methods addressing top-$N$ recommendation problems.

\citet{DBLP:conf/recsys/NingK12} propose several methods to incorporate side information with \ac{SLIM}~\cite{DBLP:conf/icdm/NingK11}. 
Among all these methods, \ac{cSLIM} generally achieves the best performance as it can well compensate sparse ratings with side information.
\citet{DBLP:conf/ijcai/ZhaoXG16,DBLP:conf/ijcai/ZhaoG17} proposed a joint model to combine self-recovery for user rating and predication from side information.
Side information is also utilized to address cold-start top-$N$ recommendation. \citet{DBLP:journals/tist/ElbadrawyK15} learn feature weights for calculating item similarities. \citet{DBLP:conf/sdm/SharmaZHK15} further improve over \cite{DBLP:journals/tist/ElbadrawyK15} by studying feature interactions.
While these methods generate the state-of-the-art performance for top-$N$ recommendation, they are all linear models, which have restricted model capacity. 

\subsection{Deep learning for hybrid recommendation}

Several authors have attempted to combine deep learning with collaborative filtering. 
\citet{DBLP:conf/wsdm/WuDZE16} utilize a denoising autoencoder to encode ratings and recover the score prediction.
\citet{DBLP:journals/nn/ZhuangZQSXH17} propose a dual-autoencoder to learn representations for both users and items. 
\citet{DBLP:conf/www/HeLZNHC17} generalize matrix factorization for collaborative filtering by a neural network.
These methods utilize user ratings only, that is, side information is not utilized.
\citet{DBLP:conf/kdd/WangWY15} propose stacked denoising autoencoders to learn item representations from side information and form a collaborative deep learning method. 
Later, \citet{DBLP:conf/cikm/LiKF15} reduce the computational cost of training by replacing stacked denoising autoencoders by a marginalized denoising autoencoder. 
Rather than manually corrupt input, variational autoencoders were later utilized for representation learning~\cite{DBLP:conf/kdd/LiS17}. 
These models achieve state-of-the-art performance among hybrid recommender systems, but they are less effective when side information is high-dimensional.
For more discussions on deep learning based recommender systems, we refer to a recent survey~\cite{DBLP:journals/corr/ZhangYS17aa}.


\section{Conclusion}\label{sec:conclusion}

In this paper, we have proposed an alternative way to feed side information to neural network so as to overcome the high-dimensionality. 
We propose \acf{cVAE}, which can be regarded as the non-linear generalization of \ac{cSLIM}. \ac{cVAE} overcomes rating sparsity by feeding both ratings and side information into the same inference network and generation network. 
To cater for the heterogeneity of information (rating and side information), we assume different sources of information to follow different distributions, which is reflected in the use of different loss function. 
As for the implementation, we introduce a parameter $\alpha$ to balance the positive samples and negative samples. 
We also introduce $\beta$ as the parameter for regularization, which controls how much the latent variable should be complied with the prior distribution.
We conduct experiments over Amazon datasets. 
The results show the superiority of \ac{cVAE} over other methods under the scenario with high-dimensional side information.

In conclusion, deep models are effective as long as the number of inputs are sufficient. 
Thus, using side information to pre-train \ac{cVAE} helps to overcome the high-dimensionality. 
A general rule-of-thumb is, regularizing \ac{cVAE} lightly during pre-train and heavily during the refinement of training.



\bibliographystyle{ACM-Reference-Format}
\bibliography{recsys}


\end{document}